\documentclass[12pt]{article}
\usepackage{graphicx} 
\usepackage{epstopdf}
\usepackage{setspace}
\usepackage{amsmath}
\usepackage[letterpaper, margin=1in]{geometry}
\usepackage{tikz}
\usetikzlibrary{snakes}
\usepackage{booktabs}
\usepackage{multirow}
\usetikzlibrary{decorations.pathreplacing,calc}
\usepackage{caption} 
\usepackage{apacite}
\usepackage{natbib}
\usepackage{adjustbox}
\usepackage{longtable}
\usepackage{subcaption}
\usepackage{threeparttable}
\usepackage{multicol}
\usepackage{float}
\usepackage{hyperref}
\usepackage{threeparttable}
\usepackage{array}
\usepackage{amsthm}
\usepackage{mathtools}
\usepackage{indentfirst}
\usepackage{pifont} 
\usepackage[justification=centering]{caption}

\theoremstyle{plain}

\theoremstyle{plain}

\title{Insuring Long-Term Care in Developing Countries: \\ The Interaction between Formal and Informal Insurance}

\author{Jiayi Wen \thanks{E-mail: wjyecon@gmail.com. School of Economics and Wang Yanan Institute of Studies in Economics (WISE), Xiamen University, Fujian, China.} \and Xiaoqing Yu  \thanks{E-mail: yxqecon@163.com. School of Economics, Xiamen University, Fujian, China} }
\date{August, 2024}

 \fontsize{12pt}{18pt}\selectfont

\begin{document}
\maketitle

\begin{abstract}

Does public insurance reduce uninsured long-term care (LTC) risks in developing countries, where informal insurance predominates? This paper exploits the rollout of LTC insurance in China around 2016 to examine the impact of public LTC insurance on healthy workers' labor supply, a critical self-insurance channel. We find that workers eligible for public LTC insurance were less likely to engage in labor work and worked fewer weeks annually following the policy change, suggesting a mitigation of uninsured risks. However, these impacts were insignificant among those with strong informal insurance coverage. Parallel changes in anticipated formal care use corroborate these findings. While our results reveal that public LTC insurance provides limited additional risk-sharing when informal insurance predominates, they also underscore its growing importance.

\end{abstract}

\newpage
     \section{Introduction}

Long-term care (LTC) risks pose a great challenge for aging societies worldwide. The World Health Organization indicates that by 2030, one-sixth of the world's population will be over 60 years old, and by 2050, this number will reach 2.1 billion. 
Significant amounts of financial risks stemming from LTC are left uninsured \citep{brown2011insuring}, and a number of studies have identified prominent self-insurance motive against LTC risks \citep{imrohorouglu2018chinese, ameriks2020long, bueren2023long}.

Governments in developed countries have spent immensely on addressing LTC risks. OECD countries spend around 1.5-2.1\% of their GDP on public LTC expenditure \citep{gruber2023long}. In contrast, public insurance against LTC risks in developing countries is significantly limited. For instance, China spent only 0.02 - 0.04\% of its GDP on public LTC expenditure \citep{glinskaya2018options}. Nevertheless, the lack of public insurance in developing countries is often compensated for by rich informal insurance via family risk-sharing \citep{chen2019hidden, lei2022longterm}.   As public LTC insurance is on a number of developing countries' agendas, the central questions of policy design are:  How much additional risk-sharing can public LTC insurance offer on top of the predominant informal insurance? How valuable is public LTC insurance for families with different levels of informal insurance coverage?

This paper examines the impact of the recent rollout of public LTC insurance in China on the labor supply of healthy older workers, an important self-insurance channel. In particular, we focus on how the impact varies by the extent of informal insurance coverage. Empirical research has primarily focused on the effects of LTC insurance on ex-post outcomes, such as caregivers' labor supply and health expenditures \citep{kim2015long,fu2017spillover, costa2018does, moura2022subsidized, coe2023family}. In contrast, evidence on the role of LTC insurance in mitigating ex-ante uninsured risk is more limited, and studies from developing countries are particularly scarce. This scarcity is largely due to the fact that few developing countries offer public LTC insurance. We leverage the recent policy change in a country where informal insurance is also prevalent.

 A direct indicator of self-insurance against LTC risks is saving. However, saving data often suffer from missing values, measurement errors, as well as reporting bias. This issue is particularly acute in developing countries, where administrative data are frequently unavailable, and research must rely heavily on survey data. To overcome this challenge, we focus on the labor supply of healthy workers before the realization of their LTC risks. Labor supply is typically well-defined, subject to fewer measurement errors, and is commonly available in most surveys. Moreover, the literature has theoretically established that labor supply is a crucial channel for insuring against shocks and smoothing consumption \citep{pijoan2006precautionary,heathcote2014consumption}. Additionally, following a growing body of research that validates the importance of labor supply in assessing the welfare gains of social insurance \citep{fadlon2019household, coyne2024household}, this paper thus focuses on labor supply as a key indicator of self-insurance.

We start by introducing the institutional background of public LTC insurance and the state of informal insurance in China. We present a simple framework to conceptualize labor supply decisions facing LTC risks and the interaction between formal and informal LTC insurance. We then employ a staggered difference-in-differences (DiD) strategy to investigate the causal effects of LTC insurance on the labor supply of healthy older workers. The analysis is based on the China Health and Retirement Longitudinal Study (CHARLS), a nationally representative dataset spanning from 2011 to 2018 across four waves, which timely covers the reform periods. Labor supply is measured by two variables separately: the labor work engagement and the number of weeks worked annually.

The empirical results suggest that public LTC insurance has meaningfully alleviated the self-insurance burden on older workers: After the rollout, both the labor work engagement rate and the annual weeks worked show disproportionately larger declines among individuals eligible for public LTC insurance compared to those who are not.  However, further analysis reveals that these effects are primarily driven by older individuals with more limited informal insurance coverage, while those with strong informal insurance experience insignificant changes in their labor supply. Specifically, we examine the effects separately based on the coresidence with adult children, a key determinant of informal care \citep{mommaerts2018coresidence}.\footnote{ \cite{stern1995estimating} and \cite{bonsang2009does} also show that geographic proximity to children is an important determinant of informal care.} The results show that the likelihood of engaging in labor work of older people who did not coreside with their adult children significantly decreased by 8 percentage points. In contrast, there was no significant change for those who coresided with their children. Similarly, the annual weeks worked for those not coresiding with their children decreased by approximately 3.8 weeks, whereas for coresidential older workers, the decline was 1.3 weeks and statistically insignificant. Although the results show that public LTC insurance offers limited additional risk-sharing for half of the current older population, it will be crucial for the other half, as well as for the additional 150 million non-coresidential older individuals projected by 2050.

We also explore the heterogeneous effects using alternative indicators of informal insurance coverage, yielding similar findings. The results show that the rollout of public LTC insurance disproportionately reduced the self-insurance burden on older individuals with fewer children and more daughters. Given that social norms in China traditionally assign the primary responsibility for old-age support to sons \citep{guo2020effects}, these findings underscore the importance of considering the interaction between economic and cultural factors when shaping public policies.

The additional analysis on anticipated formal care use supports our findings on labor supply. Overall, we find that the likelihood of anticipated formal care use increased by 3.1 percentage points following the LTC insurance rollout for individuals eligible for public insurance. However, this impact is predominantly driven by those with limited informal insurance. Individuals living independently show a 7.7-percentage-point increase in expected formal care use, while those coresiding with children exhibit an insignificant change, with a point estimate of -0.8 percentage points.

While our primary focus is on self-insurance, it is also important to examine whether public LTC insurance has crowded out informal insurance. Our results indicate that the rate of coresidence with adult children did not change significantly, nor did older individuals' expectations of receiving informal care from their adult children. This finding suggests that informal care remains preferable to formal care, consistent with our prior understanding of old-age support in China.

We also conduct several robustness analyses, including allowing for treatment effect heterogeneity based on \cite{sun2021estimating}, employing a triple-differences design, controlling for linear trends in outcomes by insurance type, and implementing placebo tests for eligibility and timing. The results remain robust across these specifications. 

This paper makes a significant contribution to understanding the ex-ante effects of public LTC insurance on self-insurance behaviors by providing causal evidence. Due to limited policy variation, most studies on the risk-sharing role of LTC insurance rely on structural analyses \citep{kopecky2014impact,barczyk2018evaluating,bueren2023long,mommaerts2024long}. Empirical studies that evaluate public LTC insurance primarily focus on ex-post outcomes - those occurring after the realization of LTC risks - such as care utilization, health status, and caregivers' labor supply \citep{kim2015long,fu2017spillover, costa2018does, moura2022subsidized, coe2023family}. A notable exception is the work by \cite{liu2023public}, which examines the effect of public insurance on consumption in China.\footnote{By focusing on labor supply, the current paper provides complementary evidences on the effect of LTC insurance on self-insurance behaviors. Furthermore, we highlight the interaction between formal and informal insurances, as explained in subsequent discussions.}

This paper also contributes to the literature by providing evidence on the risk-sharing effects of a different type of public insurance - LTC insurance. While evidence exists for public pensions mitigating longevity risks and health insurance covering acute health risks, as demonstrated by \cite{baicker2013oregon} and \cite{liu2016insuring}, research on LTC insurance remains scarce but crucial.\footnote{See \cite{banerjee2024social} for a survey on developing countries.} LTC risk is one of the largest financial risks in aging societies, prompting many governments to focus on policy designs to alleviate it \citep{brown2011insuring}.

The second major contribution of this paper is the use of a unique opportunity to assess the impact of public LTC insurance in developing countries, with a specific focus on the interaction between formal and informal insurance. A growing body of research recognizes the critical role of informal care in LTC provision \citep{barczyk2018evaluating, ko2022equilibrium, mommaerts2024long}. However, to the best of our knowledge, there is no causal evidence on how public insurance interacts with informal insurance in reducing uninsured risks. Understanding this interaction is crucial for developing countries, which currently rely heavily on informal insurance but are facing a rapid decline in its availability due to increasing migration, declining fertility rates, and rising opportunity costs of informal care.

This paper also shows that public LTC insurance has minimal impact on crowding out informal insurance, contributing to the discussion on the substitutability between formal and informal care \citep{mommaerts2018coresidence, coe2023family}.

Finally, this paper contributes to the literature on the labor supply effects of LTC insurance. While existing research predominantly focuses on the impacts on caregivers, often showing a positive effect \citep{fu2017spillover, geyer2018labor}, this paper reveals negative effects on future care recipients before the realization of LTC risks. These findings highlight an important fiscal externality that policymakers need to consider.\footnote{\cite{ai2024long} is the first work to document the negative labor supply effect of LTC insurance in China. The current paper offers new insights in three important aspects: (1)emphasizing the interaction between formal and informal insurance; (2)focusing on the role of risk-sharing of labor supply, an omitted yet critical perspective; (3)using a different identification strategy, which is discussed in the Econometric Methods section.}


\section{Background and Conceptual Framework}

\subsection{Formal and Informal Insurance in China}
Under population aging, governments are increasingly concerned about the challenges brought about by LTC risks. While developed economies spend substantially on government support, public insurance is rare among developing and middle-income countries. China stands out as a special case, as it is currently in the process of implementing such a system.

	\begin{figure}[H]
		\centering
					
			\begin{minipage}{0.99\textwidth}

    				\quad \, \includegraphics[height=2.7cm]{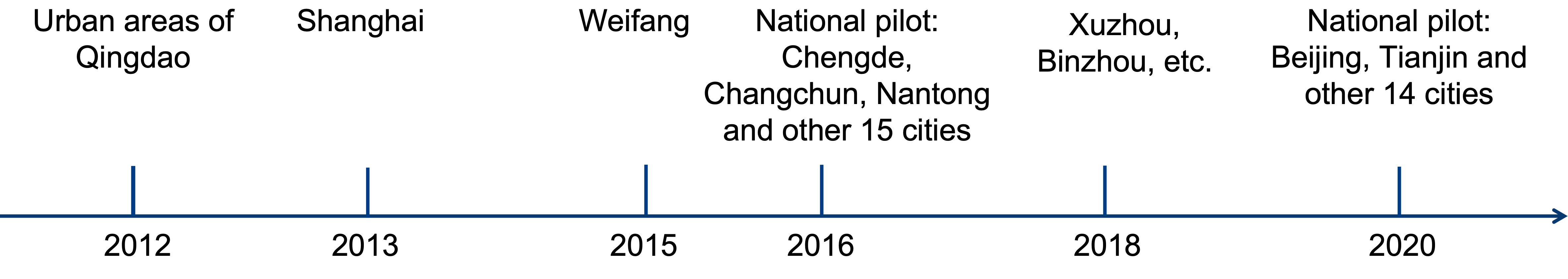} 	
						\caption{Timeline of Public LTC Insurance Rollout in China}
						\label{figure:background}   
				\tiny{\bigskip \textit{Notes:} This figure shows the timeline of the rollout of LTC insurance across pilot cities in China. \par}	
			\end{minipage}
		
	\end{figure}

China's public LTC insurance system aims to address costs of essential daily care and medical services for older people who have been severely disabled for at least six months. In 2012, the urban area of Qingdao city took the initiative of piloting public LTC insurance.  The first set of national pilot policies of LTC insurance, covering 15 cities, was officially launched in June 2016 and has since expanded its scope. As of June 2021, the LTC insurance pilot program includes 49 pilot cities, covering 134 million people.

 A key feature of China's public LTC insurance is that it is linked to the existing public health insurance system. Specifically, residents enrolled in the Urban Employee Basic Medical Insurance (UEBMI) in all pilot cities are automatically eligible for LTC insurance, while some cities extend the coverage to other types of health insurance, such as the Urban and Rural Residents Basic Medical Insurance (URRBMI). Detailed information on city-specific coverage is presented by Table \ref{table:policy0} 
 in Appendix A. In terms of insurance benefits, most 
 pilot cities have adopted a serviced-based reimbursement model, where the LTC insurance fund reimburses care costs directly to service providers. This model specifics reimbursement caps, within which the coinsurance rates typically ranging from 0\% to 30\%. 

Commercial LTC insurance is extremely uncommon in China, and older people have a strong aversion to use nursing institutions. Instead, they primarily cope with LTC risks via informal risk-sharing with adult children and other relatives. In particular, coresidence is a major living arrangement for older people in developing countries, which corresponds with timely and careful care services. Coresidential rates are substantially higher in developing countries than in developed countries. According to \cite{united2017living}, only 19.3 percent of the people aged 60 and above live with their adult children in high-income countries. In contrast, the share rises to 68.2 percent among lower-middle-income countries.
According to the census data, the rate for people aged 65 and above in China is approximately 50\%. 

However, a number of factors in developing countries are driving up the costs of informal care. First, with an increase of labor mobility, more younger individuals are working away from their parents. According to the seventh population census, the number of empty-nesters in China has reached about 120 million, accounting for 45.1\% of the country's older population. Second, rising income level significantly raises the opportunity costs for informal caregivers. The real wage rate of China in 2022 was 260\% of the one in 2008 \citep{international2022global}. 
Third, enduring reduction in fertility rates leads to fewer caregivers to share care burdens. According to \cite{chen2022path}, the average number of children is 4, 3.5, and 2.8 respectively for older individuals aged 80-84, 70-74, and 60-64 in China. As these economic and demographic changes persist, the burden on informal care are increasingly concerning, underscoring the need for alternative care solutions.

\subsection{Conceptual Framework}

\paragraph{Self-Insurance via Labor Supply} To guide our understanding of the labor supply decisions of healthy workers in anticipation of future LTC risks, we provide a simple conceptual framework by extending a standard two-period intertemporal consumption optimization problem to incorporate labor supply decisions.\footnote{For more complex quantitative models of labor supply as an insurance against future risks, see \cite{low2005self}, \cite{pijoan2006precautionary} and \cite{blundell2016consumption}.}

An individual optimizes utility over two periods $U(C_1,H)+E[U(C_2)]$, where the discount rate is set to 1 for simplicity, and labor supply $H$ is feasible only in the first period when the individual is young. LTC risks emerge in the second period, incurring LTC expenditure $D$. The budget constraint is given by $C_2= w H -C_1-D$, where $w$ is the wage rate. The individual first makes the labor supply decision by choosing $H$, then decide the optimal consumption taking $H$ as given. For a given level of $H$, the Euler equation of the intertemporal utility maximization over consumption is given by:
\begin{align}
     \frac{ \partial U (C_1, H) } {\partial C_1}  =   E \left[      \frac{ \partial U (C_2)}{\partial C_2} \right]    \label{eq:cf1}  
 \end{align}
With higher marginal utility in the second period caused by LTC expenditure $D$, the individual insure against this future risk by lowering the consumption level in the first period.\footnote{ Specifically, the higher marginal utility induced by LTC risks in the second period stems from both the intertemporal substitution and the precautionary motive. To see this, we can decompose the marginal utility in the 2nd period:
$    E[ U^{'} (C_2)   ] = U^{'} ( E[C_2] ) + \left ( E[ U^{'} (C_2)   ] - U^{'} ( E[C_2] )  \right) $.
The first term captures intertemporal substitution and the second term captures precautionary motive when the marginal utility is convex.}

Now consider the optimal labor supply condition. Denoting the optimal consumption under a given labor supply level as $C_1^{*}(H)$, it can be shown that the necessary condition for the optimal labor supply decision is:
\begin{align}
     \frac{\partial U(C_1,H)}{\partial C_1}      \frac{ \mathrm{d}  C_1^{* } (H)}{ \mathrm{d} H   }   +      \frac{\partial U(C_1,H)}{\partial H}   = &        E \left[      \frac{ \partial U (C_2)}{\partial C_2} \right]     \left[   \omega -   \frac{ \mathrm{d}  C_1^{* } (H)}{ \mathrm{d} H   }  \right]  \label{eq:cf2}
\end{align}
When $ \left[   \omega -   \frac{ \mathrm{d}  C_1^{* } (H)}{ \mathrm{d} H   }  \right] > 0  $, the right hand side is similar to Equation \ref{eq:cf1}, where the LTC expenditure $D$ leads to a higher marginal utility.\footnote{ $ \left[   \omega -   \frac{ \mathrm{d}  C_1^{* } (H)}{ \mathrm{d} H   }  \right] > 0 $ means that increasing labor supply in youth leads to higher consumption in old age, which is a general assumption. } In response, individuals in the first period can either increase savings to increase  $  \frac{\partial U(C_1,H)}{\partial C_1} $, or reduce labor supply to raise $\frac{\partial U(C_1,H)}{\partial H}$, given that  the utility function $U(C,H)$ is decreasing and concave in labor supply $H$.

\paragraph{Interaction between Formal and Informal Insurance}   We then provide heuristic discussions on implications of the interaction between formal and informal insurance against LTC risks.  We assume that  LTC expenditures $D$ can be reduced by formal or informal insurance by producing a  quantity $Y$ of LTC, whose price is normalized as 1. Therefore, $D - Y$ is the out-of-pocket expenditure that is left uninsured (self-insured).

How effective is public insurance hinges on the interaction between formal and informal care. Consider that the LTC service $Y$ is produced by a general constant elasticity of substitution (CES) technology $ Y= (\alpha K^{\rho} + (1-\alpha) L^{\rho} )^{\frac{1}{\rho} } $,
where $K$ and $L$ represent the formal and informal care respectively, and $\alpha$ is the share of formal care, also capturing its relative productivity. Under the budget constraint $r K + w  L = M$, standard CES function properties yield the optimized LTC service $Y^{*}= M/ \widetilde{p}(r,w)$, where $ \widetilde{p}(r,w)$ can be considered as a weighted average price of a unit LTC service. 

The rollout of public LTC insurance reduces the price of formal care $r$. To see how this policy change reduces the uninsured risk $D-Y^*$, the key is to understand the effect on  $\widetilde{p}(r,w)$. Standard derivation suggests the unit price $\widetilde{p}(r,w)$ has the form $ \widetilde{p}(r,w) = \left(  \alpha^\sigma r ^{1-\sigma}  +  (1 - \alpha)^\sigma w ^{1-\sigma}   \right) ^{\frac{1}{1-\sigma}} $.
Therefore, the effect of the rollout of public insurance is captured by the following first order derivative:
\begin{align}
 \frac{ \partial \widetilde{p}(r,w)}{ \partial r } =  \left(  \alpha^\sigma r ^{1-\sigma}  +  (1 - \alpha)^\sigma w ^{1-\sigma}   \right)  ^{\frac{\sigma}{1-\sigma}} \cdot \alpha^{\sigma} \cdot r ^{-\sigma}
\label{eq3}
\end{align}
This derivative is strictly positive under regular conditions, suggesting that the rollout of public insurance will lower the average price of LTC service and mitigate 
uninsured expenditure $D-Y^*$.

Meanwhile, the magnitude of the impact depends on parameters $\sigma$, $\alpha$ and $w$.\footnote{While the closed form result about how the effect varies with $w$ is clear, results about $\sigma$ and $\alpha$ is cumbersome and obscure.  We thus briefly discuss their intuition.}  An important question focused by this paper is how does the effect of public insurance vary by informal insurance coverage. The predominance of informal care in developing countries can be due to a low price $w$ or a high relative productivity of informal care $1-\alpha$. In particular, these two parameters are also likely to be heterogeneous across families. The productivity of informal care $1-\alpha$ in some families may be high, such as those with strong ties. Meanwhile, the informal care in some families may be relatively cheap if the caregivers have poor labor market opportunities. The effect of formal LTC insurance tends to be smaller for such families according to Equation \ref{eq3}.\footnote{On top of these two factors, $\sigma$ reflects the substitution between formal and informal care.  Intuitively, a large elasticity of substitution $\sigma$ in general suggests that the formal insurance tends to have a smaller effect on reducing self-insurance burden, because the informal care will also be lower as it is very elastic and substitutable in response to a lower price of formal care.}


\section{Data and Methods}
\subsection{Data}

Our analyses mainly use the CHARLS data from 2011, 2013, 2015, and 2018. The CHARLS aims to collect high-quality longitudinal data on households and individuals aged 45 and above in China, encompassing various aspects of socioeconomic status, labor market outcomes, health conditions, and intergenerational relationships. The national baseline survey was conducted in 2011, covering 150 county-level units, 450 village-level units, and approximately 17,000 individuals from about 10,000 households. Additionally, the questionnaire design of CHARLS closely follows international counterparts, including the Health and Retirement Study (HRS) from the United States and the Survey of Health, Aging, and Retirement in Europe (SHARE). Our variables are derived from three sections: the Work, Retirement, and Pension section provides information on labor market outcomes; the Healthcare and Insurance section provides data to identify eligibility for long-term care insurance; and the Family section offers insights into intergenerational relationships to identify informal insurance coverage. 

We augment the individual-level data with city-level control variables obtained from the China Urban Statistical Yearbook and the China Economic Information Center (CEIC) database.\footnote{The CEIC database combines macroeconomic data of more than 210 countries and regions, with more than 2500 data sources, such as National Bureau of Statistics, International Monetary Fund, World Bank, Organization for Economic Cooperation and Development, etc.} These city-level variables help control omitted variables that may lead to city-specific trends.
\begin{table}[H]
	\centering
	\scriptsize
	\caption{Descriptive Statistics of Main Variables}
	\begin{minipage}{0.99\textwidth}
		\tabcolsep=0.1cm
		\begin{tabular}{p{0.36\textwidth}>{\centering}p{0.12\textwidth}>{\centering}p{0.12\textwidth}>{\centering}p{0.12\textwidth}>{\centering}p{0.1\textwidth}>{\centering\arraybackslash}p{0.1\textwidth}}
			\hline
			\hline
			Variable & Observations & Mean & Std & Min & Max \\
			\hline
			Individual/household-level & & & & & \\
			\quad Labor work engagement & 7,232 & 0.689 & 0.463 & 0 & 1 \\
			\quad Annual weeks worked & 7,232 & 22.77 & 21.45 & 0 & 52.14 \\
             \quad LTC insurance & 7,232 & 0.081 & 0.273 & 0 & 1 \\
			\quad Age  & 7,232 & 58.3 & 6.4 & 45 & 69 \\
           \quad Gender& 7,232&	0.368	&0.482	&0&	1\\
			\quad Urban hukou & 7,232 & 0.267 & 0.442 & 0 & 1 \\
			\quad Receiving pension & 7,232 & 0.302 & 0.459 & 0 & 1 \\
			\quad Number of children & 7,232 & 2.17 & 1.06 & 1 & 8 \\
             \quad Coresidence & 7,232 & 0.462 & 0.499 & 0 & 1 \\
              \quad Having more sons & 7,226 & 0.663 & 0.473 & 0 & 1 \\   
			\quad Chronic disease & 7,232 & 0.704 & 0.457 & 0 & 1 \\
			\quad Self-rated Health & 7,232 & 0.796 & 0.403 & 0 & 1 \\  
			\quad Severe depression & 7,232 & 0.085 & 0.279 & 0 & 1 \\
			\quad Log of household non-financial assets & 7,232 & 7.452 & 1.449 & 1.414 & 10.61 \\
			\quad Log of household food expenditure & 7,232 & 5.193 & 0.992 & 0.639 & 7.619 \\
  			City-level   & & & & & \\ 
			\quad Hospital beds  & 7,232 & 483.7 & 178.3 & 211.8 
            & 930.9 \\ 
			\quad Old-age dependency ratio & 7,232 & 15.934 & 3.624 & 8.600 & 22.70 \\
			\quad Log of per capita GDP & 7,232 & 10.76 & 0.485 & 9.675 & 11.71 \\
			\quad Log of  fiscal expenditure& 7,232 & 8.742 & 0.578 & 7.553 & 10.66 \\
			\hline
		\end{tabular} 
  
		\label{table:des}
		\tiny{\textit{Notes:} This table presents descriptive statistics of main variables. Detailed definitions are provided in Appendix B.   \par }
	\end{minipage}
\end{table}
Several sample restrictions are imposed for subsequent empirical analyses: First, given the focus on the impacts of public insurance on ex-ante behaviors, we exclude older individuals already with ADL impairments. Second, we maintain a sample of individuals aged from five years prior to the statutory retirement age (i.e. men aged 55 and above, women aged 45 and above) to 70 years old, whose labor supply decisions are more relevant. Third, our identification strategy relies on the comparison of residents from pilot cities with and without the eligibility for public LTC insurance; therefore, we confine the sample to observations from pilot cities. Detailed discussion on this identification strategy is provided in the Econometric Methods section. Observations from the city of Qingdao are excluded to enhance comparability, as it implemented the pilot program as early as 2012. Finally, observations with missing values in main outcome and explanatory variables are also dropped. The final dataset includes 7,232 observations from 24 cities over our sample periods.\footnote{Older people who have no children, which accounts for only
1\% of the sample (76 data points) are excluded for cleaner subgroup analysis. We also perform a 1\% two-sided winsorization on household food expenditures and non-financial assets.}

	\begin{figure}[H]
		\centering
					
			\begin{minipage}{0.99\textwidth}
		 
    			\quad	\includegraphics[height=12cm]{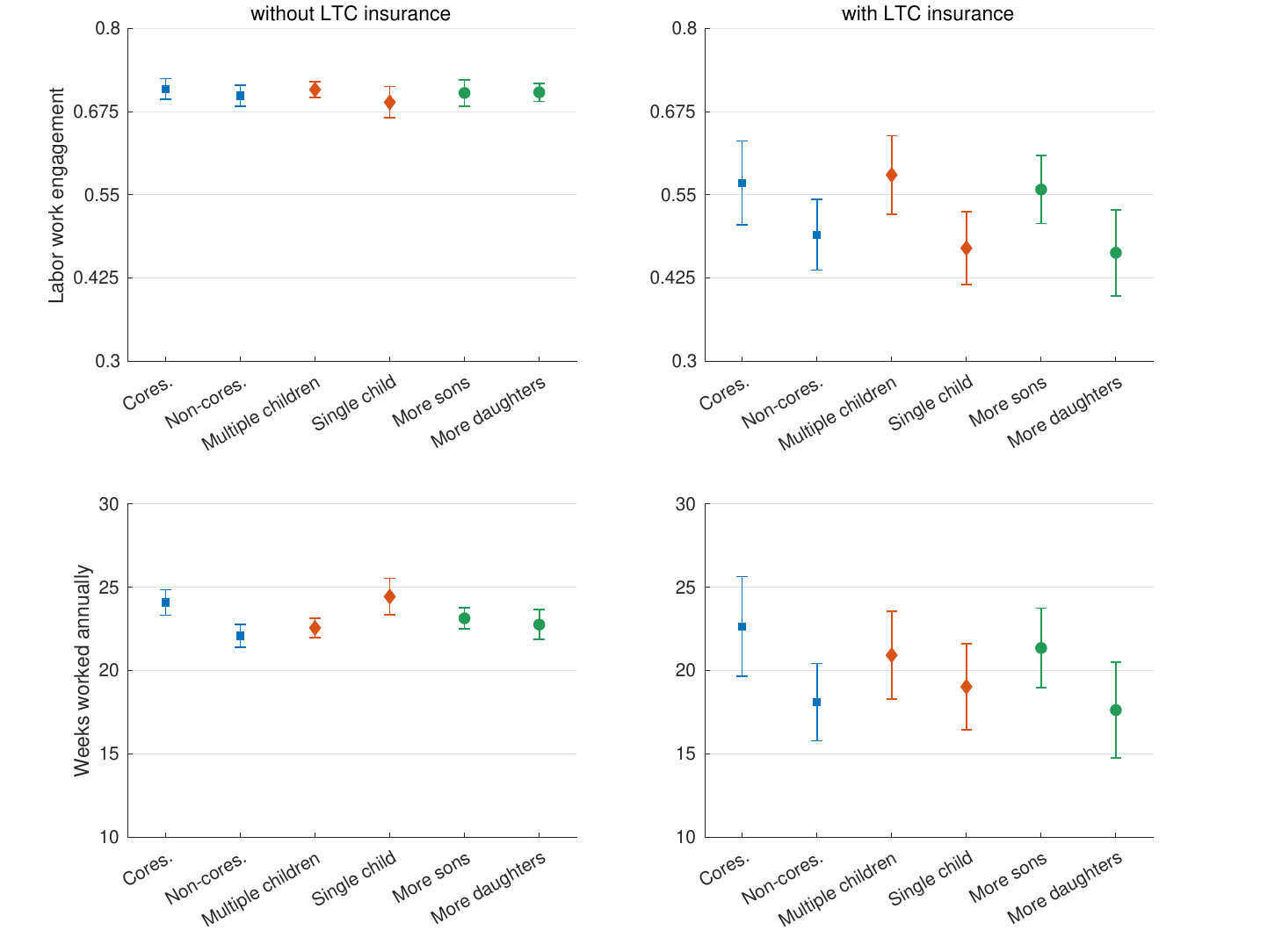} 	
						\caption{Labor Supply by Formal and Informal Insurance Coverage }
						\label{figure:subsum}   
				\tiny{\bigskip \textit{Notes:} This figure presents labor supply levels by the coverage of public LTC insurance and by informal insurance indicators. Non-coresidence, single child, and more daughters indicate more limited coverage of informal insurance. Detailed discussions are referred to Section 4.2 and 4.3. Confidence intervals are at the 95\% confidence level. \par}	
			\end{minipage}
		
	\end{figure}

Table \ref{table:des} presents descriptive statistics of the main variables, of which definitions are provided in Appendix B, Table \ref{table:vars}. The average age of the sample is approximately 58, with more females due to their longer life expectancy.  The work engagement rate stands at 69\% and the weeks worked annually are 22.8 weeks on average, with a large variation in labor supply because the sample covers ages around retirement. The coverage of public pension is low, being around 30\%, though some individuals haven't reached their pension eligible ages yet. Meanwhile, each older individual has two children on average. Nearly half of them live with their children, and about 66 percent appear to have more sons than daughters.

By the coverage of formal and informal insurance, Figure \ref{figure:subsum} 
provide motivating facts about older people's labor supply. For individuals without public LTC insurance, we find similar labor supply levels by informal insurance coverage, based on indicators of the coresidential status, number of children and whether having more sons. Interestingly, among individuals covered by public LTC insurance, older people with limited informal insurance coverage appear to had notably lower likelihood of engaging labor work and worked fewer weeks annually.  Admittedly, these gaps may be driven by a number of omitted factors, given the differences in their socioeconomic statuses. Credible evidences require analyses that exploit proper quasi-experimental variations, which are pursued by our subsequent empirical analyses.

\subsection{Econometric Methods}

\paragraph{\textit{Staggered DiD}} We use a staggered DiD design to leverage the quasi-experimental variation introduced by the rollout of public LTC insurance in China across cities during our sample period from 2013 to 2018.

The treatment group in this design includes individuals from pilot cities who have the type of public health insurance eligible for public LTC insurance, as determined based on Table \ref{table:policy0} 
in Appendix A. The control group includes individuals without such health insurance also from pilot cities.\footnote{For clarity, throughout this paper, individuals \textit{eligible} for LTC insurance refers to those having the health insurance type that qualifies for LTC insurance, including the periods before the rollout. Individuals \textit{covered} by LTC insurance refers to eligible individuals after the rollout.}
Since insurance types are endogenous, omitted variables may lead to different labor supply patterns between the treatment and control groups. The DiD design overcomes the endogeneity issue by comparing labor supply changes induced by the rollout in the treatment group relative to the control group. The core identification assumption is the conditional parallel trends assumption, which requires that the treatment and control group would have followed similar labor supply trends had the policy change not occurred. As the counterfactual trend of the treatment group is never observed, examining pre-trends prior to the treatment is a common and important test.

Specifically, the staggered DiD strategy is implemented by estimating the following econometric equation:
\begin{equation}
Y_{ict} = \beta_0 + \beta_1 LTCI_{ict} + X_{it}' \gamma + Z_{ct}' \delta + \omega_i + \mu_t + \epsilon_{it} 
\end{equation}
\( Y_{ict} \) represents the labor supply measure of individual \( i \) in city $c$ and year \( t \).  \( LTCI_{ict} \) is a dummy variable indicating the status of public LTC insurance coverage, which equals 1 if individual \( i \) is covered by LTC insurance in year \( t \), and 0 otherwise. The timing of rollout varies across pilot cities, and the type of public health insurance eligible for LTC insurance also differs. Therefore, \( LTCI_{ict} \) is determined based on whether city \( c \)  at time \( t \) has implemented the public insurance, and whether individual \( i \)  participated in the type of health insurance stipulated by the city for LTC insurance eligibility.
\( X_{it} \) is a set of time-varying control variables at the individual and household level.\footnote{These variables include age and its quadratic term, reaching the statutory retirement age, hukou status, receiving pension, number of children, chronic disease status of self and spouse, self-rated health, minor depression, moderate depression, severe depression, log of household non-financial assets and log of food expenditure. Detailed variable descriptions are provided in Table \ref{table:vars}.} \( Z_{ct} \) represents city-level time-varying 
 control variables.\footnote{ They include the number of hospital beds per thousand people, old-age dependency ratio, GDP per capita, and fiscal expenditure per capita.} \( \mu_t \) and \( \omega_i \) are time and individual fixed effects respectively.  Under the conditional parallel trend assumption,  \( \beta_1 \) captures the average treatment effect on the treated.

 It is worth discussing the strengths and weaknesses of an alternative DiD strategy that compares individuals across pilot and nonpilot cities. In our design, the control group includes residents ineligible for public LTC insurance but living also in pilot cities. Alternatively, residents of non-pilot cities with similar health insurance can also serve as the control group.\footnote{For this alternative identification strategy, see \cite{ai2024long}.}. This alternative design has the advantage of better handling omitted variable bias at the individual level. However, it requires a large set of city-specific control variables to address endogeneity issues at the city-level, such as the potential endogenous determination of pilot cities. In contrast, our strategy can better address endogeneity issues at the city level. We combine it with the rich set of detailed control variables at the individual level provided by CHARLS to ensure the conditional parallel trend assumption.\footnote{We have also tried the triple difference design. See Subsection 5.2 
 for discussions.}

\paragraph{\textit{Event Study}} To test the validity of the parallel trend assumption, we further adopt the event study strategy based on the following equation:
\begin{equation}
Y_{ict} = \sum_{p=-2}^{P} \mu_p D_{ict}^p + \sum_{q=1}^{Q} \tau_q D_{ict}^q + X_{it}' \gamma  + Z_{ct}' \beta  + \alpha_i + \lambda_t + \nu_{it} 
\end{equation}
This model compares the difference in labor supply between the treatment and control group in each period relative to the baseline period, which is set to the year before the policy change.  \( D_{ict}^q  \) is an indicator that equals 1 if the individual $i$ is from the treatment group and the period $t$ is $q$ period after the rollout. \( D_{ict}^p \)  are similar indicators for pre-rollout periods. Given that most cities rolled out the insurance program in the later years of our sample period, and that we are primarily interested in testing the pre-trends, we take $P=-7$ and $Q=3$, where we group the third, fourth, and fifth post-rollout periods due to limited observations. $ \mu_p $ captures dynamic differences in labor supply between the treatment and control groups prior to the policy rollout, while \( \tau_q \) reflects the dynamic treatment effects.

Specifically, \( \mu_p \) are used to test the differences in labor supply trends between the treatment and control groups prior to the policy change. If the trends are similar before the rollout, while a significant change occurs only in the treatment group following the policy change, this shift is most likely driven by the implementation of public LTC insurance.


\section{Empirical Results}

\subsection{Overall Effects on Labor Supply}

We first examining the overall impact of public LTC insurance on healthy older workers' labor supply. Based on the staggered DiD design, we find that individuals eligible for public LTC significantly reduced their labor supply after the policy rollout. Their likelihood of engaging in any labor work decreased by 9.2 
percentage points relative to those ineligible. Meanwhile, they worked 3.4 fewer weeks annually. Both estimates are significant at the 99\% confidence level.  

According to the conceptual framework, public insurance lowers the marginal utility of future consumption by insuring future LTC risks, leading individuals to increase their leisure and lower their labor supply at the current stage. Our finding indicates nontrivial ex-ante impacts of public insurance in alleviating self-insurance burden against future LTC risks.

	\begin{figure}[H]
		\centering
			\begin{minipage}{0.99\textwidth}
   \hspace{0.5cm} \includegraphics[height=7cm]{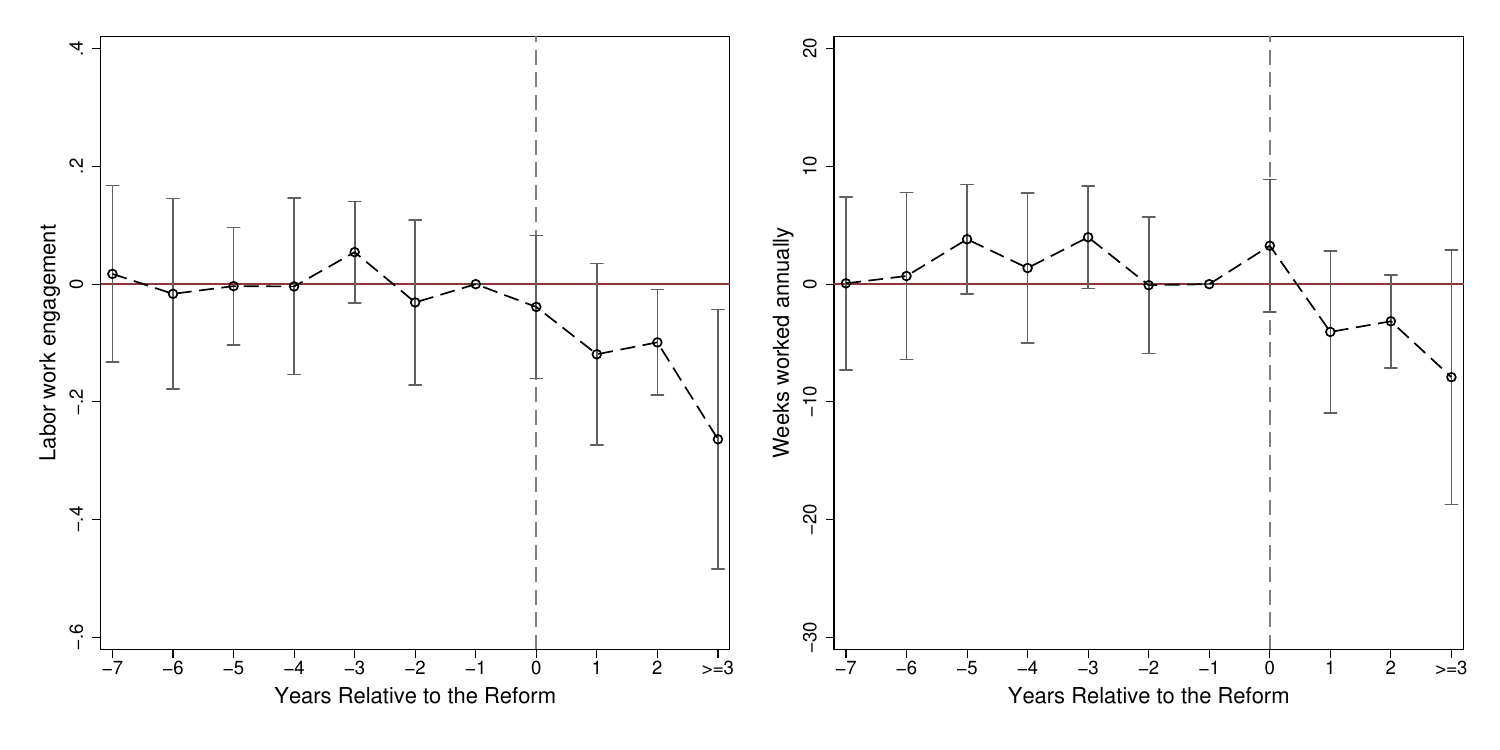} 	
						\caption{ Difference in Labor Supply Trends between  Individuals \\ Eligible and Ineligible for LTC Insurance  }
						\label{figure:ptrends}   
				\tiny{\bigskip \textit{Notes:} This figure illustrates the changes in labor supply of individuals eligible for LTC insurance relative to those ineligible. Confidence intervals are at the 95\% confidence level. \par}	
			\end{minipage}	
	\end{figure}

Figure \ref{figure:ptrends} presents the difference in labor supply trends between the treatment and control group estimated by the event study design. It is clear that up to seven years before the rollout of LTC insurance, individuals eligible and ineligible for LTC insurance exhibited similar trends in both labor work engagement rates and annual weeks worked, whereas these similar trends began to diverge since the rollout of public insurance. This event study result supports the parallel trends assumption for our DiD design.

\subsection{Interaction with Informal Insurance}

With limited access to public and commercial insurance in developing countries, families play a crucial role in risk-sharing. Can public insurance mitigate uninsured LTC risks in the presence of informal insurance?

According to our conceptual framework, the impact of public LTC insurance depends on the relative quality and cost of formal and informal care. If the quality of informal care is sufficiently high or its cost is sufficiently low, formal insurance will offer limited additional risk protection for those already covered by informal arrangements. As a result, families with high-quality or affordable informal care will likely exhibit small responses to public insurance reforms. The impact also depends on the elasticity of substitution between formal and informal insurance. For instance, if they are perfectly substitutable, individuals originally relying on informal insurance may simply replace it with formal insurance after the reform, leaving uninsured risks unchanged. Ultimately, the effectiveness of public insurance in reducing uninsured LTC risks is an empirical question.

Therefore, we empirically examine the role of informal insurance in the rollout of public LTC insurance. Given that coresidence has been a central focus in the literature on informal care  \citep{johar2014does, fu2017spillover, mommaerts2018coresidence, coe2023family}, we use it as a primary indicator of informal insurance coverage, based on whether older individuals live with their adult children. Coresidence typically implies lower costs of informal care provision compared to cases where adult children live far away. It also tends to indicate higher quality of care, as the frequency of intergenerational contact is often a measure of old-age support.

Table \ref{table:core} reveals that the overall effects of public LTC insurance on labor supply are primarily driven by older individuals with limited informal insurance coverage. Specifically, following the rollout of public insurance, there has been a notable 8-percentage-point decrease in labor work engagement rates among older individuals who live apart from their adult children. In contrast, those living with their children experienced a smaller, statistically insignificant decrease of 5 percentage points. Additionally, non-coresidential older individuals worked 3.8 weeks less annually following the policy change, whereas coresidential older individuals reduced their annual work weeks by only 1.3 weeks, which is also statistically insignificant.

 \begin{table}[H]
 \centering
 \scriptsize
\caption{Effects of LTC Insurance by Coresidence with Adult Children}
 \begin{minipage}{0.99\textwidth}
   \tabcolsep=0.005cm\begin{tabular}{p{0.25\textwidth}>{\centering}p{0.2\textwidth}>
   {\centering}p{0.175\textwidth}>
   {\centering}p{0.2\textwidth}>  
{\centering\arraybackslash}p{0.175\textwidth}}
\hline
\hline
    & \multicolumn{2}{c}{Coresidence} & \multicolumn{2}{c}{Non-Coresidence}  \\
      \cmidrule(lr){2-3}  \cmidrule(lr){4-5} 
       & Engagement & Weeks & Engagement & Weeks \\
      \hline
      LTC insuarance& -0.050 & -1.272 & -0.079** & -3.783** \\
          & (0.047) & (2.174) & (0.036) & (1.710) \\
     Observations & 3,338 & 3,338 & 3,894 & 3,894 \\
      R-squared & 0.025 & 0.031 & 0.029 & 0.043 \\
               Dep. Mean  & 0.699 & 23.982 & 0.680 & 21.722 \\
  \hline
\end{tabular} 
  \label{table:core}
  {\tiny \textit{Notes:} This table presents the baseline estimates using the staggered DiD approach. Engagement refers to labor work engagement, and Weeks refers to weeks worked annually. *** p$<$0.01, ** p$<$0.05, * p$<$0.1.  All control variables are included. Standard errors clustered at the individual level are in parentheses. \par } 
 \end{minipage}  
\end{table}

These findings suggest that while public insurance reduces overall uninsured LTC risks, it appears to offer limited additional risk-sharing for individuals already covered by informal insurance from their extended family. Nevertheless, alongside the current older population, population aging is projected to result in another 102.6 million older individuals living independently by 2050, assuming the current 46\% co-residential rate. Moreover, living arrangements may change. If the coresidential rate declines to the level seen in high-income countries (approximately 19\%), an additional 50.73 million older people will urgently need public LTC insurance.

\subsection{Alternative Indicators of Informal Insurance Coverage} 
We also explore alternative indicators of informal insurance coverage based on the number of children and the proportion of sons among all adult children. The number of children is straightforward, while the proportion of sons captures social norms about old-age support in China and other Asian countries.\footnote{ \cite{guo2020effects} points out that in China and many other developing countries, sons are considered better providers of old-age support than daughters, also citing evidence from \cite{ebenstein2010missing} and \cite{huang2017love}.}  
Specifically, our subsequent analyses are separately conducted based on whether older individuals have multiple children and whether they have more sons.

 The results are presented in Table \ref{table:other}. We find that the labor supply impacts of LTC insurance are primarily driven by older individuals with only one child and more daughters, who significantly reduce their labor work engagement rates by 13.4 and 13.7 percentage points, respectively. In contrast, those with more children and more sons show notably smaller responses, with reductions of 4.8 and 7 percentage points, respectively. Similarly, the changes in annual work weeks also exhibit significant differences. Among those with only one child and more daughters, annual work weeks decreased substantially by 5.5 and 5.9 weeks. Conversely, the declines were 1.1 and 2.1 weeks among those with more children and more sons, which were statistically insignificant. 
 
 \begin{table}[H]
 \centering
 \tiny
\caption{Effects of LTC Insurance by Alternative Informal Insurance Indicators}
 \begin{minipage}{0.99\textwidth}
   \tabcolsep=0.05cm\begin{tabular}{p{0.15\textwidth}>{\centering}p{0.1\textwidth}>
   {\centering}p{0.1\textwidth}>
   {\centering}p{0.1\textwidth}>  
   {\centering}p{0.1\textwidth}>    
   {\centering}p{0.1\textwidth}>
   {\centering}p{0.1\textwidth}>
   {\centering}p{0.1\textwidth}>
{\centering\arraybackslash}p{0.1\textwidth}}
\hline
\hline
     & \multicolumn{4}{c}{Multiple children} & \multicolumn{4}{c}{More sons} \\
        & \multicolumn{2}{c}{Yes} & \multicolumn{2}{c}{No}  & \multicolumn{2}{c}{Yes} & \multicolumn{2}{c}{No} \\
     \cmidrule(lr){2-3} \cmidrule(lr){4-5} \cmidrule(lr){6-7} \cmidrule(lr){8-9}  
                 & Engagement & Weeks & 
                 Engagement & Weeks & Engagement & Weeks & Engagement & Weeks \\
      \hline
      LTC insurance  & -0.048 & -1.088 & -0.134*** & -5.487*** & -0.070** & -2.070 & -0.137*** & -5.887*** \\
        & (0.036) & (1.696) & (0.041) & (1.855) & (0.033) & (1.539) & (0.043) & (1.972) \\

            Observations & 5,309 & 5,309 & 1,923 & 1,923 & 4,788 & 4,788 & 2,438 & 2,438 \\
R-squared & 0.019 & 0.027 & 0.071 & 0.099 & 0.018 & 0.031 & 0.062 & 0.063 \\
            Dep. Mean & 0.702 & 22.477 & 0.653 & 23.563 & 0.693 & 23.006 & 0.681 & 22.282 \\
  \hline
\end{tabular}  
  \label{table:other}
  {\tiny \textit{Notes:} This table presents the estimates based on alternative indicators of informal insurance coverage. Engagement refers to labor work engagement, and Weeks refers to weeks worked annually.  *** p$<$0.01, ** p$<$0.05, * p$<$0.1.  All control variables are included. Standard errors clustered at the individual level are in parentheses. \par } 
 \end{minipage}  
\end{table}

\subsection{Expected Use of Formal LTC} 

Is the reduced labor supply really driven by the rollout of public LTC insurance? To offer reassurance, we further examine older individuals' anticipated usage of formal care. If we observe no similar effects on expected formal care usage, we might concern that the previous findings on labor supply could be due to factors other than the reform.

Specifically, we exploit information solicited by the questions: ``\textit{(1) Suppose that in the future, you needed help with basic daily activities like eating or dressing. Do you have relatives or friends (besides your spouse/partner) that would be willing and able to help you over a long period of time?}''; and ``\textit{(2) What is the relationship to you of that person or those persons?}''.\footnote{These questions should be interpreted with caution, as the first question specifies relatives or friends. However, the second question offers a wide range of options that are not only relatives or friends. The wording is also consistent across waves, and we focus on variations across waves.} We code the dependent variable as 1 if the answers include ``\textit{a paid helper (such as nanny)}'' or  ``\textit{Volunteer, Employee(s) of facility, and Community}'', and 0 otherwise.

\begin{table}[H]
 \centering
 \scriptsize
\caption{Effects of LTC Insurance on Anticipated Use of Formal and Informal LTC}
 \begin{minipage}{0.99\textwidth}
   \tabcolsep=0.01cm\begin{tabular}{p{0.15\textwidth}>{\centering}p{0.12\textwidth}>
{\centering}p{0.15\textwidth}>
  {\centering}p{0.15\textwidth}>
{\centering}p{0.12\textwidth}>
{\centering}p{0.15\textwidth}>
{\centering\arraybackslash}p{0.15\textwidth}}
\hline
\hline
& \multicolumn{3}{c}{Formal Care}   & \multicolumn{3}{c}{Informal Care} \\    \cmidrule(lr){2-4} \cmidrule(lr){5-7}
& Overall & Coresidence & Non-coresidence  & Overall & Coresidence & Non-coresidence  \\
\hline 
LTC insurance  & 0.031** & -0.008 & 0.077*** & 0.021 & -0.003 & 0.010 \\
          & (0.012) & (0.017) & (0.022) & (0.032) & (0.051) & (0.050) \\
Observations & 6,939 & 3,195 & 3,744 & 6,939 & 3,195 & 3,744 \\
R-squared & 0.019 & 0.030 & 0.041 & 0.032 & 0.036 & 0.053 \\
 Dep. Mean & 0.016 & 0.012 & 0.019 & 0.684 & 0.713 & 0.660 \\
  \hline
\end{tabular} 
  \label{table:exp}
  \tiny{\textit{Notes:} This table presents the estimates based on the dependent variable of whether the individual expects to use formal or informal LTC in the future.  *** p$<$0.01, ** p$<$0.05, * p$<$0.1.  All control variables are included. Standard errors clustered at the individual level are in parentheses. \par }
 \end{minipage}  
\end{table}

As shown in Table \ref{table:exp}, on average, public LTC insurance has increased older individuals' anticipated use of formal care by 3.1 percentage points when future daily activities are restricted. Remarkably, we similarly find that there was no impact among older individuals who coresided with their children, whereas the expected formal care use among those living independently increased by as much as 7.7 percentage points. This result indicates that the risk protection of public insurance is indeed perceived by older individuals,  being consistent with the reduced self-insurance burden revealed by their actual labor supply responses.

\subsection{Impacts on Informal Insurance}

\label{subs43}

Why does public insurance provide limited additional risk protection among older people who are already covered by informal insurance? Understanding the reason behind is crucial for future policy design, and two alternative explanations emerge. First, formal and informal care may be highly substitutable, and informal insurance coverage is elastic in response to better risk protection from formal insurance.  This mechanism seems consistent with the literature that identifies large spillover effects on caregivers' labor supply.\footnote{ For example, see \cite{fu2017spillover}. However, note that this spillover effect on caregivers primarily happens after the realization of risks. The ex-ante substitution between formal and informal insurance against LTC risks can be different.}. Alternatively, the quality and price of informal care may dominate, resulting in a minimal response to policy changes of formal care.

Therefore, we further explore the effects of public LTC insurance rollout on informal insurance. As revealed by Table \ref{table:exp}, there is neither overall effect nor coresidential heterogeneity in the anticipation of future care provided by adult children. In terms of actual behaviors, we also find no discernible change in living arrangements among older people eligible for public LTC insurance compared to those ineligible. The statistically insignificant effect of LTC insurance rollout on coresidential rate is 0.017 with a standard error 0.026.

These findings align with the second explanation and our prior knowledge that older people are more willing to use informal care. They are also consistent with \cite{mommaerts2018coresidence}, who found that LTC subsidy policies in the U.S.(Medicaid) have insignificant effects on the likelihood of coresidence among single older individuals younger than 80.\footnote{She find significant effects only for people older than 80.}


\section{Robustness}
 In this section, we conduct several robustness tests to provide reassuring evidence for our main findings.

\subsection{Treatment Effect Heterogeneity}
Our main identification exploits labor supply variations of the treatment group relative to the control group, before and after the rollout of LTC insurance. As pointed out by recent developments in theoretical DiD literature \citep{goodman2021difference}, using already treated observations as the control group may lead to biased estimates under the staggered design with heterogeneous treatment effects. To rule out this ``forbidden'' comparison, we replicate our baseline regressions based on  the interaction-weighted estimator proposed by \cite{sun2021estimating}. This new specification uses individuals who have never been treated as the control group, excluding observations that have already been covered by LTC insurance.

Table \ref{table:sun} presents our new results, revealing that LTC insurance has a negative effect on the labor work engagement rate, with an estimated decrease of 12.3 percentage points. There is also a marked reduction in the number of weeks worked annually by 4.3 weeks. These estimates are even larger than ones of the main specification. When examining by living arrangements, coresidential older individuals show minor and insignificant changes in both measures of labor supply, whereas those living independently reduce their labor supply significantly, particularly in terms of annual weeks worked.

\begin{table}[H]
 \centering
 \scriptsize
\caption{Effects of LTC Insurance Estimated Based on \cite{sun2021estimating}} 
 \begin{minipage}{0.99\textwidth}
   \tabcolsep=0.02cm\begin{tabular}{p{0.21\textwidth}>{\centering}p{0.15\textwidth}>
   {\centering}p{0.12\textwidth}>
   {\centering}p{0.12\textwidth}>
   {\centering}p{0.13\textwidth}>
   {\centering}p{0.12\textwidth}>
    {\centering\arraybackslash}p{0.13\textwidth}}
\hline
\hline
& \multicolumn{2}{c}{Overall} &  \multicolumn {2}{c}{Coresidence} & \multicolumn{2}{c}{Non-Coresidence}\\
      \cmidrule(lr){2-3}  \cmidrule(lr){4-5}   \cmidrule(lr){6-7} 
& Engagement & Weeks & Engagement & Weeks& Engagement & Weeks  \\
   \hline
LTC insurance & -0.123*** &  -4.273** &  -0.100 & -2.914 & -0.123**  &    -7.752***  \\
     & (0.042)    & (  1.671)  & ( 0.066)  & (   2.498) & (0.059)  & (  2.875)  \\
\hline
Observations &  6,404 &  6,404 & 2,434  & 2,434 & 2,905  &   2,905 \\
R-squared &  0.648 & 0.616 & 0.671 & 0.650 & 0.687  & 0.655  \\
Dep. mean& 0.695 &  22.834 &  0.700 &  24.130 & 0.691 &  21.810   \\
\hline
\end{tabular} 
  \label{table:sun}
  \tiny{\textit{Notes:} This table presents the estimates using the IW estimator proposed by \cite{sun2021estimating}.  Engagement refers to labor work engagement, and Weeks refers to weeks worked annually.  *** p$<$0.01, ** p$<$0.05, * p$<$0.1.  All control variables are included. Standard errors clustered at the individual level are in parentheses. \par }
 \end{minipage}  
\end{table}

\subsection{Triple Difference}

Our main identification strategy exploits variations among residents from pilot cities, with and without the types of public health insurance eligible for LTC insurance. In principle, residents from non-pilot cities could also serve as the control group. Nevertheless, as explained in the Econometric Methods section, we have a variety of individual-level variables but limited city-level covariates to control for. Therefore, the conditional parallel trend assumption is more likely to hold when comparing different individuals within similar cities.

As a robustness check, we further adopt a triple difference (DDD) strategy to use the sample of non-pilot cities. The advantage of the DDD design is that, even if there were differential trends of labor supply between the eligible and ineligible individuals in the pilot cities, these trends could be controlled for by subtracting trends estimated on the non-pilot cities, provided that these trends are similar across pilot and non-pilot cities.\footnote{We prefer the DiD over the DDD design for three reasons. First, while the DDD design significantly expands the sample size, all additional data points accrue to the control group, exacerbating the imbalance between the treated and control groups in our context. Second, it seems less imperative to control for differential trends between the eligible and ineligible residents in pilot cities, given that the pre-trends are parallel. Third, the DiD design is more transparent and interpretable.}

Specifically, we estimate the following econometric model:
\begin{align*}
Y_{ict} = & \beta_0 + \beta_1  Elig_{ict} \times Pilot_{c} \times Post_{ct} +\beta_2 Elig_{ict} \times Pilot_{c} \\ 
&  +\beta_3 Pilot_{c} \times Post_{ct} + X_{it}' \gamma + Z_{ct}' \delta + \omega_i + \mu_t + \epsilon_{it} 
\end{align*}
\(Pilot_{c}\) is a dummy variable indicating whether a city is a pilot city. \(Elig_{ict}\) indicates whether an individual has the public health insurance that qualifies the LTC insurance as stipulated by the living city.  \(Post_{ct}\)  is a dummy variable indicating that the period is after the policy change of city c.

Table \ref{table:ddd} demonstrates the results estimated by the DDD design and the findings remain similar to our main estimates by DiD.

\begin{table}[H]
 \centering
 \scriptsize
\caption{Effects of LTC Insurance Estimated by Triple Difference }
\begin{minipage}{0.99\textwidth}
   \tabcolsep=0.02cm\begin{tabular}{p{0.21\textwidth}>{\centering}p{0.15\textwidth}>
   {\centering}p{0.12\textwidth}>
   {\centering}p{0.12\textwidth}>
   {\centering}p{0.13\textwidth}>
   {\centering}p{0.12\textwidth}>
    {\centering\arraybackslash}p{0.13\textwidth}}
\hline
\hline
& \multicolumn{2}{c}{Overall} &  \multicolumn {2}{c}{Coresidence} & \multicolumn{2}{c}{Non-Coresidence}\\
      \cmidrule(lr){2-3}  \cmidrule(lr){4-5}   \cmidrule(lr){6-7} 
& Engagement & Weeks & Engagement & Weeks& Engagement & Weeks  \\
   \hline
LTC insurance & -0.089*** & -3.715*** & -0.069 & -2.711 & -0.082**  & -3.660** \\
    & (0.026)   & (1.223)   & (0.047) & (2.277) & (0.037)  & (1.710) \\
\hline
Observations & 31,676 & 31,676 & 16,223 & 16,223 & 15,453  & 15,453 \\
R-squared & 0.016   & 0.021   & 0.014 & 0.018   & 0.019     & 0.026 \\
Dep. mean  & 0.692 & 22.638 & 0.695 & 23.152 & 0.688  & 22.098 \\
\hline
\end{tabular} 
  \label{table:ddd}
  \tiny{\textit{Notes:} This table presents the estimates using the the triple difference estimator.  Engagement refers to labor work engagement, and Weeks refers to weeks worked annually.  *** p$<$0.01, ** p$<$0.05, * p$<$0.1.  All control variables are included. Standard errors clustered at the individual level are in parentheses. \par }
 \end{minipage}  
\end{table}

\subsection{Group-Specific Linear Trends }

Our analyses compare older individuals with and without public health insurance types eligible for LTC insurance. The DiD strategy accounts for difference in levels while assuming that the outcomes would have followed similar trends had the policy change not occurred. Figure \ref{figure:ptrends} shows no significant difference in their outcome trends prior to the rollout of LTC insurance, making it reasonable to expect these similar trends to continue after the reform, unless some unknown events occurred concurrently.


One may still worry about trends specific to older individuals eligible for LTC insurance. For instance, those with UEBMI 
often have access to employee public pensions, whose benefits have been increased annually. This adjustment may reduce the likelihood of labor work engagement.\footnote{This concern is mostly relevant if the benefit adjustment occurred in the same year as the rollout of LTC insurance, which is not the case in reality.}

We thus further account for the linear trends in dependent variables specific to the treatment group by including an interaction term between the UEBMI indicator and a linear function of time.   Table \ref{table:lnt} shows results similar to our baseline regressions, indicating that the main findings are not driven by group-specific linear trends.

\begin{table}[H]
 \centering
 \scriptsize
\caption{ Effects of LTC Insurance Controlling for Group-specific Linear Trends }
\begin{minipage}{0.99\textwidth}
   \tabcolsep=0.02cm\begin{tabular}{p{0.28\textwidth}>{\centering}p{0.14\textwidth}>
   {\centering}p{0.12\textwidth}>
   {\centering}p{0.11\textwidth}>
   {\centering}p{0.11\textwidth}>
   {\centering}p{0.11\textwidth}>
    {\centering\arraybackslash}p{0.11\textwidth}}
\hline
\hline
& \multicolumn{2}{c}{Overall} &  \multicolumn {2}{c}{Coresidence} & \multicolumn{2}{c}{Non-Coresidence}\\
      \cmidrule(lr){2-3}  \cmidrule(lr){4-5}   \cmidrule(lr){6-7} 
& Engagement & Weeks & Engagement & Weeks& Engagement & Weeks  \\
   \hline
LTC insurance & -0.085*** & -2.393 & 0.016& 1.312  & -0.090* & -3.539* \\
          & (0.033)   & (1.502) & (0.056)  & (2.638)  & (0.047)& (2.135) \\
UEBMI*Years & -0.002   & -0.355  & -0.024*  & -0.928 & 0.004    & -0.083 \\
          & (0.007)  & (0.334) & (0.012) & (0.564) & (0.011)  & (0.504) \\
\hline
Observations & 7,232 & 7,232 & 3,338  & 3,338 & 3,894 & 3,894 \\
R-squared & 0.023 & 0.037 & 0.028 & 0.033 & 0.029  & 0.043 \\
Dep. mean   & 0.689 & 22.765 & 0.699 & 23.982 & 0.680 & 21.722  \\
\hline
\end{tabular} 
  \label{table:lnt}
  \tiny{\textit{Notes: }This table presents the estimates controlling for group-specific linear trends. UEBMI*Year is the interaction between a dummy variable of UEBMI and the year. Engagement refers to labor work engagement, and Weeks refers to weeks worked annually.  *** p$<$0.01, ** p$<$0.05, * p$<$0.1.  All control variables are included. Standard errors clustered at the individual level are in parentheses. \par }
 \end{minipage}  
\end{table}

\subsection{Falsification Tests }
Following convention, we also perform falsification tests to investigate whether the estimated effects indeed come from the rollout of public LTC insurance. To this end, we set a falsified year of the rollout, or a falsified type of public health insurance eligible for LTC insurance.

Since nearly all policy changes were implemented after 2013 and before 2018, as shown in Table \ref{table:policy0}, we select 2012 as the falsified year, as there were no actual pilot programs launched in this year in our sample. As shown in Table \ref{table:pl2012}, no significant effects are detected for either labor work engagement or annual weeks worked.

Additionally, we consider a hypothetical scenario where the URRBMI is the type of public health insurance that qualifies for LTC insurance in each city. The timing of policy changes remains aligned with the reality.  As shown in Table \ref{table:pl2012}, there is once again no significant effect.\footnote{While such falsification tests seem conventional, a caveat is that what these falsified regressions really estimate are unclear in econometric theory. For instance, by setting a falsified treatment group, observations from the actual treatment group now become the control group and the estimated effects are obscure. }

\begin{table}[H]
 \centering
 \scriptsize
\caption{Effects of LTC Insurance Based on Falsified Policies }
\begin{minipage}{0.99\textwidth}
   \tabcolsep=0.03cm\begin{tabular}{p{0.22\textwidth}>{\centering}p{0.13\textwidth}>
   {\centering}p{0.12\textwidth}>
   {\centering}p{0.13\textwidth}>
   {\centering}p{0.12\textwidth}>
   {\centering}p{0.13\textwidth}>
    {\centering\arraybackslash}p{0.12\textwidth}}
\hline
\hline
& \multicolumn{2}{c}{Overall} &  \multicolumn {2}{c}{Coresidence} & \multicolumn{2}{c}{Non-Coresidence}\\
      \cmidrule(lr){2-3}  \cmidrule(lr){4-5}   \cmidrule(lr){6-7} 
& Engagement & Weeks & Engagement & Weeks& Engagement & Weeks  \\
   \hline
\multicolumn{7}{l}{\textit{A. Falsified Timing of Program Rollout}}   \\
LTC Insurance & -0.020 & -1.053 & -0.028& -1.312 & -0.017  & -1.595 \\
          & (0.025) & (1.228) & (0.041)& (1.979) & (0.043)  & (2.075) \\
\hline
\multicolumn{7}{l}{\textit{B. Falsified Eligibility for LTC Insurance}}   \\
LTC Insurance & -0.013 & -0.103 & -0.058 & -2.116 & -0.019  & -0.857 \\
          & (0.027) & (1.240) & (0.047)& (2.398) & (0.039)  & (1.685) \\
\hline
Observations & 7,232 & 7,232 & 3,338  & 3,338& 3,894 & 3,894 \\
R-squared & 0.020 & 0.035 & 0.024 & 0.031& 0.027  & 0.041 \\
Dep. mean & 0.689 & 22.765& 0.699 & 23.982 & 0.680 & 21.722  \\
\hline
\end{tabular} 
  \label{table:pl2012}
  
  \tiny{\textit{Notes:} This table presents the estimates using falsified treatments with 2012 as the year of rollout.  *** p$<$0.01, ** p$<$0.05, * p$<$0.1.  All control variables are included. Standard errors clustered at the individual level are in parentheses. \par }
 \end{minipage}  
\end{table}

\subsection{Only UEBMI}

The basic type of public health insurance that qualifies for LTC insurance is UEBMI, while a handful of cities extend the eligibility to other types of health insurance, such as URRBMI. Our main analyses consider the eligibility for LTC insurance specific to each city,  as detailed in Table \ref{table:policy0} in Appendix A. One concern is that cities extending LTC insurance coverage may have stronger fiscal capacity or other unobserved factors that could also impact older workers' labor supply. 

To provide further robust evidence, we exclude cities with extended coverage and restrict our sample to those granting eligibility only based on UEBMI. The results are presented in Table \ref{table:cid}. The effects of public LTC insurance remain similar to our main results, as do the effects across different informal insurance coverage.
\begin{table}[H]
 \centering
 \scriptsize
\caption{Effects of LTC Insurance Restricted to Eligibility Only with UEBMI}
 \begin{minipage}{0.99\textwidth}
   \tabcolsep=0.03cm\begin{tabular}{p{0.22\textwidth}>{\centering}p{0.13\textwidth}>
   {\centering}p{0.12\textwidth}>
   {\centering}p{0.13\textwidth}>
   {\centering}p{0.12\textwidth}>
   {\centering}p{0.13\textwidth}>
    {\centering\arraybackslash}p{0.12\textwidth}}
\hline
\hline
& \multicolumn{2}{c}{Overall} &  \multicolumn {2}{c}{Coresidence} & \multicolumn{2}{c}{Non-Coresidence}\\
      \cmidrule(lr){2-3}  \cmidrule(lr){4-5}   \cmidrule(lr){6-7} 
& Engagement & Weeks & Engagement & Weeks& Engagement & Weeks  \\
   \hline
 LTCI  & -0.107*** & -3.680** & -0.097& -2.809 & -0.106**  & -4.068** \\
     & (0.036)   & (1.694)  & (0.075) & (3.437) & (0.047)  & (2.044) \\
\hline
Observations & 5,898 & 5,898 & 2,671 & 2,671 & 3,227  & 3,227 \\
R-squared & 0.022 & 0.029 & 0.035 & 0.030 & 0.030  & 0.040 \\
Dep. mean & 0.720 &  23.078 & 0.726 &  24.126 & 0.715  &  22.210  \\
\hline
\end{tabular} 
  \label{table:cid}
  
  \tiny{\textit{Notes:} This table presents the estimates considering only UEMBI for LTCI eligibility. *** p$<$0.01, ** p$<$0.05, * p$<$0.1.  All control variables are included. Standard errors clustered at the individual level are in parentheses. \par }
 \end{minipage}  
\end{table}

\subsection{Sample with Invariant Eligibility }
In our baseline regressions, individuals are considered treated if they had an eligible health insurance by the time of program rollout. This specification helps maintain a relatively large sample size, but there is concern that individuals' eligibility may vary over time.

To address this concern, we exclude individuals with varying eligibility for LTC insurance. As Table \ref{table:ltci-eli} indicates, regression coefficients from these new analyses do not deviate from the baseline estimates. Our main results are robust in a sample with invariant eligibility for LTC insurance.

\begin{table}[H]
 \centering
 \scriptsize
\caption{Effects of LTC Insurance Excluding Sample with Changes in LTCI eligibility }
\begin{minipage}{0.99\textwidth}
   \tabcolsep=0.03cm\begin{tabular}{p{0.22\textwidth}>{\centering}p{0.13\textwidth}>
   {\centering}p{0.12\textwidth}>
   {\centering}p{0.13\textwidth}>
   {\centering}p{0.12\textwidth}>
   {\centering}p{0.13\textwidth}>
    {\centering\arraybackslash}p{0.12\textwidth}}
\hline
\hline
& \multicolumn{2}{c}{Overall} &  \multicolumn {2}{c}{Coresidence} & \multicolumn{2}{c}{Non-Coresidence}\\
      \cmidrule(lr){2-3}  \cmidrule(lr){4-5}   \cmidrule(lr){6-7} 
& Engagement & Weeks & Engagement & Weeks& Engagement & Weeks  \\
   \hline
LTCI& -0.084*** & -3.268*** & -0.058 & -1.109 & -0.067*  & -3.715** \\
          & (0.027)   & (1.261)   & (0.048)& (2.287)  & (0.038) & (1.784) \\
\hline
Observations & 7,050 & 7,050 & 3,255 & 3,255 & 3,795  & 3,795 \\
R-squared & 0.021 & 0.035 & 0.027 & 0.031 & 0.027  & 0.042 \\
Dep. mean  & 0.692 & 22.769 & 0.700 & 23.958 & 0.684  & 21.750 \\
\hline
\end{tabular} 
  \label{table:ltci-eli}
  
  \tiny{\textit{Notes:} This table presents the estimates after removing the sample with changes in LTC insurance eligibility.  *** p$<$0.01, ** p$<$0.05, * p$<$0.1.  All control variables are included. Standard errors clustered at the individual level are in parentheses. \par }
 \end{minipage}  
\end{table}

\subsection{Samples with Invariant Coresidence }
Individuals' coresidence status may vary over time.  Our main specifications refrain from arbitrarily classifying individuals with varying coresidence as either coresidential or non-coresidential. Instead, the subgroup analyses separate observations of the same individual based on their coresidence status in each period. Importantly, our results in Subsection \ref{subs43} also show no \textit{systematic} changes in coresidential status driven by the policy change.

Nevertheless, in this subsection, we restrict our sample to individuals who have never changed their coresidence status. As revealed in Table \ref{table:corchange}, the sample size becomes substantially smaller as it includes only those who have maintained the same living arrangement across all waves. Consequently, the coefficients become insignificant. However, it is noteworthy that the magnitudes and the heterogeneous effects remain consistent with our previous findings.

\begin{table}[H]
 \centering
 \scriptsize
\caption{Effects of LTC Insurance Excluding Sample with Changes in Coresidence Status}
 \begin{minipage}{0.99\textwidth}
   \tabcolsep=0.03cm\begin{tabular}{p{0.22\textwidth}>{\centering}p{0.13\textwidth}>
   {\centering}p{0.12\textwidth}>
   {\centering}p{0.13\textwidth}>
   {\centering}p{0.12\textwidth}>
   {\centering}p{0.13\textwidth}>
    {\centering\arraybackslash}p{0.12\textwidth}}
\hline
\hline
& \multicolumn{2}{c}{Overall} &  \multicolumn {2}{c}{Coresidence} & \multicolumn{2}{c}{Non-Coresidence}\\
      \cmidrule(lr){2-3}  \cmidrule(lr){4-5}   \cmidrule(lr){6-7} 
& Engagement & Weeks & Engagement & Weeks& Engagement & Weeks  \\
   \hline
LTCI & -0.047 & -2.209 & -0.028 & -1.490 & -0.072  & -2.768 \\
          & (0.035) & (1.661) & (0.053) & (2.496)  & (0.047) & (2.236) \\
\hline
Observations & 3,652 & 3,652 & 1,686 & 1,686 & 1,966  & 1,966 \\
R-squared & 0.029 & 0.034 & 0.035 & 0.041 & 0.045  & 0.041 \\
Dep. mean  & 0.674 & 22.735 & 0.694 & 24.694 & 0.657  & 21.055 \\
\hline
\end{tabular} 
  \label{table:corchange}
  
  \tiny{\textit{Notes:} This table presents the estimates after removing the sample with changes in coresidence status.   *** p$<$0.01, ** p$<$0.05, * p$<$0.1.  All control variables are included. Standard errors clustered at the individual level are in parentheses. \par }
 \end{minipage}  
\end{table}


\section{Conclusion}
LTC risks impose significant financial burdens on aging societies. While developed countries have more established public insurance systems to protect older populations against these risks, in developing countries, informal insurance plays a crucial role in risk-sharing and consumption smoothing. However, in response to emerging demographic and economic changes, an increasing number of governments in developing economies are considering public support. The key question is whether the provision of public LTC insurance will meaningfully reduce uninsured risks, given the predominance of informal insurance.

This study presents causal evidence on the effect of public LTC insurance in mitigating uninsured risks across families with different levels of informal insurance coverage, leveraging the recent launch of a public LTC insurance pilot program in China. We find that the reform has a notable impact on lowering the labor supply of healthcare workers, as reflected by both the labor engagement rate and the number of weeks worked annually, indicating that public insurance has meaningfully reduced uninsured LTC risks.

However, our further analyses reveal that these impacts are mainly driven by families with weak coverage of informal insurance. For families with coresidence, more children, and more sons, the labor supply impacts are small and insignificant. Further analyses also reveal that the launch of public insurance has increased older people's anticipation of utilizing formal care in the future, but once again, the effect disappears among those coresiding with their adult children. Additionally, by examining the impacts on coresidential status and the anticipated care provided by adult children, we do not find evidence that public insurance has crowded out informal insurance, suggesting that using informal care is more preferable.

While our findings indicate that public LTC insurance provides little additional risk protection for families already insured by informal means, several factors - such as the rising mobility of younger generations, declining fertility, and increasing market wages - suggest a tendency for reduced coverage of informal insurance over time. The significant impacts observed among older individuals with limited informal insurance coverage - whose number is projected to increase by hundreds of millions - underscore the urgency and importance of designing public protection.

\newpage
\bibliographystyle{apalike}
\bibliography{reference}

\begin{thebibliography}{}

\bibitem[Ai et~al., 2024]{ai2024long}
Ai, J., Feng, J., and Zhang, X. (2024).
\newblock Long-term care insurance coverage and labor force participation of older people: Evidence from china.
\newblock {\em China Economic Review}, page 102192.

\bibitem[Ameriks et~al., 2020]{ameriks2020long}
Ameriks, J., Briggs, J., Caplin, A., Shapiro, M.~D., and Tonetti, C. (2020).
\newblock Long-term-care utility and late-in-life saving.
\newblock {\em Journal of Political Economy}, 128(6):2375--2451.

\bibitem[Baicker et~al., 2013]{baicker2013oregon}
Baicker, K., Taubman, S.~L., Allen, H.~L., Bernstein, M., Gruber, J.~H., Newhouse, J.~P., Schneider, E.~C., Wright, B.~J., Zaslavsky, A.~M., and Finkelstein, A.~N. (2013).
\newblock The oregon experiment—effects of medicaid on clinical outcomes.
\newblock {\em New England Journal of Medicine}, 368(18):1713--1722.

\bibitem[Banerjee et~al., 2024]{banerjee2024social}
Banerjee, A., Hanna, R., Olken, B., and Lisker, D. (2024).
\newblock Social protection in the developing world.
\newblock {\em Journal of Economic Literature}.

\bibitem[Barczyk and Kredler, 2018]{barczyk2018evaluating}
Barczyk, D. and Kredler, M. (2018).
\newblock Evaluating long-term-care policy options, taking the family seriously.
\newblock {\em The Review of Economic Studies}, 85(2):766--809.

\bibitem[Blundell et~al., 2016]{blundell2016consumption}
Blundell, R., Pistaferri, L., and Saporta-Eksten, I. (2016).
\newblock Consumption inequality and family labor supply.
\newblock {\em American Economic Review}, 106(2):387--435.

\bibitem[Bonsang, 2009]{bonsang2009does}
Bonsang, E. (2009).
\newblock Does informal care from children to their elderly parents substitute for formal care in europe?
\newblock {\em Journal of health economics}, 28(1):143--154.

\bibitem[Brown and Finkelstein, 2011]{brown2011insuring}
Brown, J.~R. and Finkelstein, A. (2011).
\newblock Insuring long-term care in the united states.
\newblock {\em Journal of Economic Perspectives}, 25(4):119--142.

\bibitem[Bueren, 2023]{bueren2023long}
Bueren, J. (2023).
\newblock Long-term care needs and savings in retirement.
\newblock {\em Review of Economic Dynamics}, 49:201--224.

\bibitem[Chen et~al., 2019]{chen2019hidden}
Chen, L., Fan, H., and Chu, L. (2019).
\newblock The hidden cost of informal care: An empirical study on female caregivers' subjective well-being.
\newblock {\em Social Science \& Medicine}, 224:85--93.

\bibitem[Chen et~al., 2022]{chen2022path}
Chen, X., Giles, J., Yao, Y., Yip, W., Meng, Q., Berkman, L., Chen, H., Chen, X., Feng, J., Feng, Z., et~al. (2022).
\newblock The path to healthy ageing in china: a peking university--lancet commission.
\newblock {\em The Lancet}, 400(10367):1967--2006.

\bibitem[Coe et~al., 2023]{coe2023family}
Coe, N.~B., Goda, G.~S., and Van~Houtven, C.~H. (2023).
\newblock Family spillovers and long-term care insurance.
\newblock {\em Journal of Health Economics}, 90:102781.

\bibitem[Costa-Font et~al., 2018]{costa2018does}
Costa-Font, J., Jimenez-Martin, S., and Vilaplana, C. (2018).
\newblock Does long-term care subsidization reduce hospital admissions and utilization?
\newblock {\em Journal of health economics}, 58:43--66.

\bibitem[Coyne et~al., 2024]{coyne2024household}
Coyne, D., Fadlon, I., Ramnath, S.~P., and Tong, P.~K. (2024).
\newblock Household labor supply and the value of social security survivors benefits.
\newblock {\em American Economic Review}, 114(5):1248--1280.

\bibitem[Ebenstein, 2010]{ebenstein2010missing}
Ebenstein, A. (2010).
\newblock The “missing girls” of china and the unintended consequences of the one child policy.
\newblock {\em Journal of Human resources}, 45(1):87--115.

\bibitem[Fadlon and Nielsen, 2019]{fadlon2019household}
Fadlon, I. and Nielsen, T.~H. (2019).
\newblock Household labor supply and the gains from social insurance.
\newblock {\em Journal of Public Economics}, 171:18--28.

\bibitem[Fu et~al., 2017]{fu2017spillover}
Fu, R., Noguchi, H., Kawamura, A., Takahashi, H., and Tamiya, N. (2017).
\newblock Spillover effect of japanese long-term care insurance as an employment promotion policy for family caregivers.
\newblock {\em Journal of health economics}, 56:103--112.

\bibitem[Geyer and Korfhage, 2018]{geyer2018labor}
Geyer, J. and Korfhage, T. (2018).
\newblock Labor supply effects of long-term care reform in germany.
\newblock {\em Health Economics}, 27(9):1328--1339.

\bibitem[Glinskaya and Feng, 2018]{glinskaya2018options}
Glinskaya, E. and Feng, Z. (2018).
\newblock {\em Options for aged care in China: Building an efficient and sustainable aged care system}.
\newblock World Bank Publications.

\bibitem[Goodman-Bacon, 2021]{goodman2021difference}
Goodman-Bacon, A. (2021).
\newblock Difference-in-differences with variation in treatment timing.
\newblock {\em Journal of Econometrics}, 225(2):254--277.

\bibitem[Gruber et~al., 2023]{gruber2023long}
Gruber, J., McGarry, K.~M., and Hanzel, C. (2023).
\newblock Long-term care around the world.
\newblock Technical report, National Bureau of Economic Research.

\bibitem[Guo and Zhang, 2020]{guo2020effects}
Guo, R. and Zhang, J. (2020).
\newblock The effects of children's gender composition on filial piety and old-age support.
\newblock {\em The Economic Journal}, 130(632):2497--2525.

\bibitem[Heathcote et~al., 2014]{heathcote2014consumption}
Heathcote, J., Storesletten, K., and Violante, G.~L. (2014).
\newblock Consumption and labor supply with partial insurance: An analytical framework.
\newblock {\em American Economic Review}, 104(7):2075--2126.

\bibitem[Huang et~al., 2017]{huang2017love}
Huang, F., Jin, G.~Z., and Xu, L.~C. (2017).
\newblock Love, money, and parental goods: Does parental matchmaking matter?
\newblock {\em Journal of Comparative Economics}, 45(2):224--245.

\bibitem[{\.I}mrohoro{\u{g}}lu and Zhao, 2018]{imrohorouglu2018chinese}
{\.I}mrohoro{\u{g}}lu, A. and Zhao, K. (2018).
\newblock The chinese saving rate: Long-term care risks, family insurance, and demographics.
\newblock {\em Journal of Monetary Economics}, 96:33--52.

\bibitem[{International Labour Organization}, 2022]{international2022global}
{International Labour Organization} (2022).
\newblock Global wage report 2022-23: The impact of inflation and covid-19 on wages and purchasing power.

\bibitem[Johar and Maruyama, 2014]{johar2014does}
Johar, M. and Maruyama, S. (2014).
\newblock Does coresidence improve an elderly parent's health?
\newblock {\em Journal of Applied Econometrics}, 29(6):965--983.

\bibitem[Kim and Lim, 2015]{kim2015long}
Kim, H.~B. and Lim, W. (2015).
\newblock Long-term care insurance, informal care, and medical expenditures.
\newblock {\em Journal of public economics}, 125:128--142.

\bibitem[Ko, 2022]{ko2022equilibrium}
Ko, A. (2022).
\newblock An equilibrium analysis of the long-term care insurance market.
\newblock {\em The Review of Economic Studies}, 89(4):1993--2025.

\bibitem[Kopecky and Koreshkova, 2014]{kopecky2014impact}
Kopecky, K.~A. and Koreshkova, T. (2014).
\newblock The impact of medical and nursing home expenses on savings.
\newblock {\em American Economic Journal: Macroeconomics}, 6(3):29--72.

\bibitem[Lei et~al., 2022]{lei2022longterm}
Lei, X., Bai, C., Hong, J., and Liu, H. (2022).
\newblock Long-term care insurance and the well-being of older adults and their families: Evidence from china.
\newblock {\em Social Science \& Medicine}, 296:114745.

\bibitem[Liu et~al., 2023]{liu2023public}
Liu, H., Ma, J., and Zhao, L. (2023).
\newblock Public long-term care insurance and consumption of elderly households: evidence from china.
\newblock {\em Journal of Health Economics}, 90:102759.

\bibitem[Liu, 2016]{liu2016insuring}
Liu, K. (2016).
\newblock Insuring against health shocks: Health insurance and household choices.
\newblock {\em Journal of health economics}, 46:16--32.

\bibitem[Low, 2005]{low2005self}
Low, H.~W. (2005).
\newblock Self-insurance in a life-cycle model of labour supply and savings.
\newblock {\em Review of Economic Dynamics}, 8(4):945--975.

\bibitem[Mommaerts, 2018]{mommaerts2018coresidence}
Mommaerts, C. (2018).
\newblock Are coresidence and nursing homes substitutes? evidence from medicaid spend-down provisions.
\newblock {\em Journal of Health Economics}, 59:125--138.

\bibitem[Mommaerts, 2024]{mommaerts2024long}
Mommaerts, C. (2024).
\newblock Long-term care insurance and the family.
\newblock {\em accpeted at Journal of Political Economy}.

\bibitem[Moura, 2022]{moura2022subsidized}
Moura, A. (2022).
\newblock Do subsidized nursing homes and home care teams reduce hospital bed-blocking? evidence from portugal.
\newblock {\em Journal of Health Economics}, 84:102640.

\bibitem[Pijoan-Mas, 2006]{pijoan2006precautionary}
Pijoan-Mas, J. (2006).
\newblock Precautionary savings or working longer hours?
\newblock {\em Review of Economic dynamics}, 9(2):326--352.

\bibitem[Stern, 1995]{stern1995estimating}
Stern, S. (1995).
\newblock Estimating family long-term care decisions in the presence of endogenous child characteristics.
\newblock {\em Journal of human Resources}, pages 551--580.

\bibitem[Sun and Abraham, 2021]{sun2021estimating}
Sun, L. and Abraham, S. (2021).
\newblock Estimating dynamic treatment effects in event studies with heterogeneous treatment effects.
\newblock {\em Journal of Econometrics}, 225(Themed Issue: Treatment Effect 1):175--199.

\bibitem[{United Nations}, 2017]{united2017living}
{United Nations} (2017).
\newblock Living arrangements of older persons: A report on an expanded international dataset.

\end{thebibliography}

\newpage

\appendix

\section*{Appendix A: LTC Insurance Pilot Rollouts }
\begin{table}[H]
 \centering
 \scriptsize
\caption{Long-Term Care Insurance Policy Pilot from 2012 to 2018} 
 \begin{minipage}{0.9\textwidth}
   \tabcolsep=0.02cm\begin{tabular}{p{0.15\textwidth}>
{\centering}p{0.5\textwidth}>
 {\centering\arraybackslash}p{0.33\textwidth}}
\hline
Year & City & Eligibility\\
\midrule
2012 & Qingdao Urban Area & UEBMI, URRBMI \\
2013 & Shanghai& UEBMI \\
2014 & Dongying & UEBMI \\
2015 & Qingdao (Expanded Coverage) & UEBMI, URBMI, URRBMI \\
2015 & Weifang & UEBMI \\
2016 & Chengde, Jinan & UEBMI \\
2016 & Songyuan, Jilin, Changchun & UEBMI, URRBMI \\
2016 & Nantong, Jingmen & UEBMI, URBMI, URRBMI \\
2017 & Anqing, Shangrao, Xuzhou, Chengdu, Linyi, Guangzhou, Liaocheng, Tai'an, Linfen, Qiqihar, Chongqing, Ningbo & UEBMI \\
2017 &Meihekou, Tonghua, Baishan & UEBMI, URRBMI \\
2017 &Shanghai (Expanded Coverage), Hangzhou, Shihezi, Jiaxing, Suzhou & UEBMI, URBMI, URRBMI \\
2018 & Binzhou, Heze, Zibo, Zaozhuang, Yantai, Weihai, Rizhao & UEBMI \\
2018 & Xuzhou (Expanded Coverage), Beijing & UEBMI, URBMI, URRBMI \\
\hline
\end{tabular} \\
  \label{table:policy0}

  \tiny{Notes: UEBMI stands for Urban Employee Basic Medical Insurance; URBMI represents Urban Resident Basic Medical Insurance; and URRBMI stands for Urban and Rural Resident Basic Medical Insurance.}
 \end{minipage}  
\end{table}
Some pilot cities are not surveyed by CHARLS. Excluding the urban areas of Qingdao, our sample covers 24 pilot cities, which are Shanghai\footnote{In 2013, Shanghai initiated a pilot medical care program to reimburse nursing costs for elderly urban workers, effectively serving as a precursor to the formal long-term care insurance later implemented, which is often overlooked in existing related research.}, Weifang, Qingdao, Chengde, Jinan, Jilin City, Jingmen, Shangrao, Anqing, Xuzhou, Chengdu, Guangzhou, Linyi, Liaocheng, Linfen, Qiqihar, Chongqing, Ningbo, Hangzhou, Jiaxing, Suzhou, Binzhou, Zaozhuang, Weihai, and Dezhou.

\section*{Appendix B: Variable Definitions } 
\begin{table}[H]
	\centering
	\scriptsize
	\caption{Definitions of Main Variables}
	\begin{minipage}{0.99\textwidth}
		\tabcolsep=0.01cm
		\begin{tabular}{p{0.27\textwidth}>{\centering\arraybackslash}p{0.72\textwidth}}
			\hline
			\hline
			VARIABLES     &  DESCRIPTION    \\
			\hline
			\textbf{Individual/household-level} &      \\
			\quad Labor work engagement & 1 if the respondent is currently engaged in any labor work, and 0 otherwise. \\
			\quad Weeks worked annually & Respondent’s number of weeks worked in the past year as reported. \\
			\quad LTC insurance & 1 if the respondent is covered by public long-term care insurance, and 0 otherwise. \\
			\quad Age & Respondent’s age \\
			\quad Gender & 1 for male, 0 for female \\
			\quad Urban hukou & 1 if the respondent has an urban household registration, and 0 otherwise. \\
			\quad Receiving pension & 1 if currently receiving a pension, and 0 otherwise. \\
			\quad Number of children & Respondent’s number of children \\
		\quad	Coresidence & 1 if the respondent coresides with any children and 0 otherwise     \\
   	\quad 		Having more sons &    1 if the respondent has a greater number of sons than daughters and 0 otherwise   \\
			\quad Chronic disease & 1 if respondent has a chronic disease, and 0 otherwise. \\
			\quad Spouse chronic disease & 1 if respondent’s spouse has a chronic disease, and 0 otherwise. \\
			\quad Self-rated health & 1 for self-reported health status good, very good, or excellent, and 0 for fair or poor \\
			\quad Depression &  Minor, moderate and severe three dummies based on respondent’s CES-D scale score.  \\   
			\quad Log of  non-financial assets & 
            Natural logarithm of the amount of household non-financial assets (excluding housing). \\
			\quad Log of food expenditure &Natural logarithm of the household's food expenditure from the previous week. \\
			\textbf{City-level} &      \\   
			\quad Hospital beds & Number of hospital beds per thousand people in the city over the past year. \\
			\quad Old-age dependency ratio & Ratio of the aged population to the working-age population, per 100 people. \\
			\quad Log of per capita GDP &  Logarithm of per capita gross domestic product for the past year in the city. \\
			\quad Log of fiscal expenditure & Logarithm of  per capita fiscal expenditure for the past year in the city. \\
			\hline
		\end{tabular} \\
		\label{table:vars}
		\tiny{\textit{Notes:}  This table presents the definitions of main variables used in our analyses.  \par }
	\end{minipage}
\end{table}%

\end{document}